# Understanding Fairness and its Impact on Quality of Service in IEEE 802.11

Michael Bredel, Markus Fidler
Multimedia Communications Lab, TU Darmstadt

*Abstract*— The Distributed Coordination Function (DCF) aims at fair and efficient medium access in IEEE 802.11. In face of its success, it is remarkable that there is little consensus on the actual degree of fairness achieved, particularly bearing its impact on quality of service in mind. In this paper we provide an accurate model for the fairness of the DCF. Given $M$ greedy stations we assume fairness if a tagged station contributes a share of $1/M$ to the overall number of packets transmitted. We derive the probability distribution of fairness deviations and support our analytical results by an extensive set of measurements. We find a closed-form expression for the improvement of long-term over short-term fairness. Regarding the random countdown values we quantify the significance of their distribution whereas we discover that fairness is largely insensitive to the distribution parameters. Based on our findings we view the DCF as emulating an ideal fair queuing system to quantify the deviations from a fair rate allocation. We deduce a stochastic service curve model for the DCF to predict packet delays in IEEE 802.11. We show how a station can estimate its fair bandwidth share from passive measurements of its traffic arrivals and departures.

## I. INTRODUCTION

The Distributed Coordination Function (DCF) specifies a randomized access procedure for the shared wireless medium in IEEE 802.11. The target is to divide resources fairly among an unknown number of stations while minimizing access delays and maximizing overall throughput. Originating from the basic ALOHA access scheme significant progress has been made regarding throughput and stability of today's Medium Access Control (MAC) protocols [1], [2]. The issue of per-flow fairness is however, still under debate and different studies do not agree in their conclusions, e.g. [3], [4]. An important aspect of fair scheduling is the attainable quality of service. While long-term fairness ensures a certain average throughput the issue of short-term fairness has tremendous impact on individual packet delays [3], [5], [6].

Centralized fair scheduling algorithms on the other hand are very well understood today. The pioneering Generalized Processor Sharing (GPS) model [7] assumes a weighted resource allocation that is perfectly fair on any time scale. To date, a variety of packet-by-packet implementations exist that emulate GPS closely, such as Weighted Fair Queuing. Distributed emulations are proposed in [8], [9], [10] with the aim to implement fair scheduling in the DCF. Models for fair packet scheduling are derived e.g. in [11], [12], [13], [14]. These models define error terms that specify the worst-case deviation of a packet scheduler from an ideal GPS system.

Moreover, the GPS model and the calculus for network delay [15] gave rise to the important concept of deterministic service curve [7] that is the foundation of today's network calculus [16], [17]. Recently, significant progress has been made towards the formulation of stochastic service curves, see [18], [19], [20] and references therein. These models are used in [21], [20], [22] to derive service curve representations of wireless links with a focus on channel outages that are due to fading and interference. Modeling random medium access is, however, an open challenge.

In this paper we analyze the fairness of the DCF and the impacts on quality of service at a single radio channel. Our contributions are as follows. First, we derive closed-form solutions for the conditional distribution $P[K = k|l]$ that a contending station transmits $k$ packets given a tagged station transmits $l$ packets within the same time interval. This characterization of fairness turns out to be comprehensive and versatile, e.g. the well-known fairness index by Jain [23] follows readily. We substantiate our analytical findings using an extensive baseline set measurements that we conducted in a shielded and unechoic room and of OmNet++ simulations. Second, we view the DCF as emulating the GPS policy. We formulate a recursive model for packet departure times coined DCF clock that is subject to well-defined random error terms. Based on the distribution of packet inter-transmissions we derive a stochastic service curve model for the DCF. Finally, we show how a station can obtain reliable estimates of the fair rate from passive measurements of its arrivals and departures.

The remainder of this paper is structured as follows. In Sect. II we discuss related work on fairness in IEEE 802.11. In Sect. III we elaborate on our controlled evaluation environment. We compare our set of baseline measurements to OmNet++ simulations as well as to related studies. In Sect. IV we develop a model of the DCF and derive closed-form expressions for the fairness. In Sect. V we view the DCF as emulating GPS and derive the DCF clock and after that a DCF service curve model in the network calculus. In Sect. VI we show how the fair share under the DCF can be estimated from passive measurements. Sect. VII provides brief conclusions.

## II. BACKGROUND AND RELATED WORK

In this section we introduce the basic functionality of the IEEE 802.11 DCF and discuss related work on its performance. The DCF seeks to achieve per-packet fairness [4], [24] that is also the subject this work. Besides, the notion of airtime fairness [25], [26], [27] is important in IEEE 802.11. It depends, however, on further parameters such as packet sizes and rate-adaptation. Numerous works, e.g. [8], [9], [10],

[28], seek to modify the DCF to improve fairness issues in wireless networks. Yet, we find that not even the degree of per-packet fairness that is achieved by the standard DCF is well-understood and precisely modeled.

## A. DCF Mode of Operation

The IEEE 802.11 standard [29] specifies the DCF mode of operation. Here, we only consider the case of greedy stations that persistently contend for the wireless medium.

We consider a tagged station that transmits a packet and after the so-called Short Inter-Frame Space (SIFS) receives the corresponding acknowledgement. Before sending the next packet the station generates a *uniformly distributed backoff value* in $[0, 1, \ldots, w-1]$ where $w$ is the contention window. Whenever the medium is continuously idle for the duration of a Distributed Inter-Frame Space (DIFS) the station starts respectively continues its countdown procedure and decrements its backoff value by one each slot time. The countdown is paused if the medium gets busy again and it is resumed after each successful DIFS waiting. As soon as the backoff value reaches zero the station sends its packet. If several stations perform the countdown procedure simultaneously the one with the smallest backoff value starts its transmission first.

If two or more stations finish the countdown procedure at the same time their transmissions cause a collision. In this case the respective stations perform an *exponential backoff*, i.e they double their current contention window $w$ at most up to $w_{max}$. They return to $w_{min}$ in case of a successful transmission.

## B. Related Work on Fairness

Achieving good throughput while taking fair medium access into account is a major goal of MAC protocols. This target gave rise to significant research, for a survey see e.g. [30]. Using a distributed scheduling approach such as the DCF it is, however, still a challenge.

A seminal performance analysis of the saturation throughput of the IEEE 802.11 DCF is provided in [2]. Based on a two-dimensional Markov model the distribution of the contention window is derived. The approach is simplified in [5] using a one-dimensional Markov chain and approximated using backoff values sampled from an exponential distribution with a mean related to the collision probability in [31]. Channel access delays are derived e.g. in [5], [6], [32].

One of the first papers that point at the importance of fairness in wireless networks is [1]. The authors investigate fair bandwidth allocation in the presence of hidden terminals and introduce the RTS/CTS message exchange. Fairness is frequently quantified using Jain's fairness index [23]. Given a number of samples $K$ the index is defined by their first and second moment as

$$f = \frac{\mathsf{E}[K]^2}{\mathsf{E}[K^2]}. \quad (1)$$

Jain's fairness index is in the interval $[0, 1]$ where larger values indicate better fairness.

An empirical study of fairness in WaveLANs is provided in [3]. Using a sliding window the throughput of individual stations is measured for different window sizes to compute Jain's fairness index for short and long time-scales. The finding is that WaveLANs are long-term fair but unfair on short time-scales. Similar results are reported for IEEE 802.11 by a number of subsequent papers that compute Jain's fairness index from simulation respectively measurement data [8], [33], [10]. Based on a Markov model the study [34] supports short-term unfairness. In contrast the authors of [4], [24] report long as well as short-term fairness.

The argument of short-term fairness is supported in [4] using an analytical model that introduces a new indicator of fairness. To this end, the packet inter-transmissions of a contending station are counted before a tagged station transmits a single packet. Considering independent and identically distributed (i.i.d.) backoff values $b_i(j)$ and two stations indexed $i = 1, 2$, the first station sends exactly $k$ packets before the second station completes its countdown if $\sum_{j=1}^{k} b_1(j) \leq b_2(1)$ and $\sum_{j=1}^{k+1} b_1(j) > b_2(1)$. The backoff values are modeled as continuous and uniformly distributed in $[0, w-1]$, hence collisions are not considered in [4]. Furthermore, synchronized stations are assumed that start their countdown procedures at the same time. The sums of uniform random variables are expressed using the Irwin-Hall distribution yielding the conditional probability that station one transmits $k$ packets given station two transmits a single packet

$$\mathsf{P}[K = k|1] = \frac{k+1}{(k+2)!}.$$

The paper [4] reports a good match of the model with empirical data and concludes short-term fairness for two stations. The closed-form result is not extended to more than two stations. The authors provide, however, measurement results in [24] which indicate that fairness decreases as the number of contending stations increases.

A number of studies [8], [10], [9] seek to develop improvements of the DCF that resemble fair scheduling algorithms. Also, the bandwidth estimation study [35] relates fairness of the DCF to the fair share achieved under GPS [7] and reports a good match for long-term fairness. Here, the fair share $f$ is defined as

$$f : \sum_{i} \min(\rho_i, f) = C \quad (2)$$

where $\rho_i$ is the source data rate of station $i$ and $C$ is the channel capacity. Despite the efforts to implement fair medium access, models that quantify deviations of the DCF from the fair share on different time scales remain an open challenge.

## III. EMPIRICAL FAIRNESS EVALUATION

We conduct an empirical evaluation of the fairness achieved by the DCF that acts as a baseline for our model-based analysis in Sect. IV. We perform extensive experiments with two and more contending stations using an IEEE 802.11 testbed in a specific shielded and unechoic measurement chamber. Hence, we can assume that the wireless medium is free of interference from external sources that do not belong to our testbed. Furthermore, we present results from OmNet++ [36] simulations of the same topology and setup for comparison.

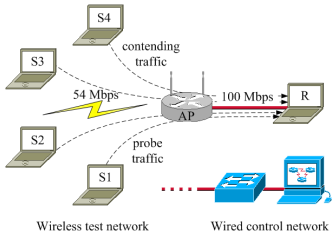
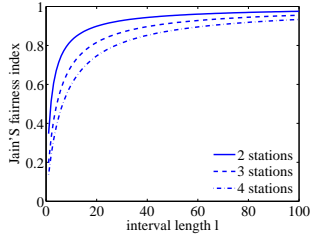

Fig. 1. Wireless testbed setup

Fig. 2. Jain's fairness index

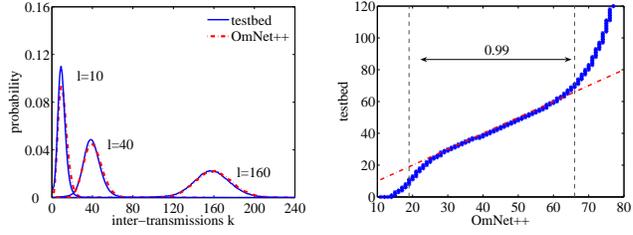

(a) pmf of inter-transmissions

(b) corresponding q-q plot for $l = 40$

Fig. 3. OmNet++ simulation results agree closely with the measurement data. The q-q plot matches well for 0.99 of the samples, but brings out deviations at the tail, showing additional unfairness in the testbed compared to the simulator.

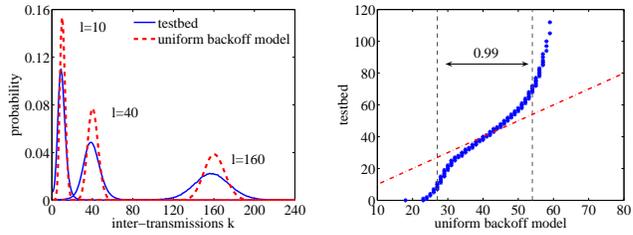

(a) pmf of inter-transmissions

(b) corresponding q-q plot for $l = 40$

Fig. 4. Numerical results from an extension of the model in [4]. The uniform backoff random variables overestimate fairness. The q-q plot not only differs at the tail but also its slope midway does not become one. Moreover, the 0.99 interval narrows by a factor of 1.7 and wrongly indicates good fairness.

Our testbed shown in Fig. 1 consists of four wireless stations (S1 to S4) that run the rude/crude traffic generator [37] to contend for the medium and send data to a receiver (R). The stations are connected to the access point (AP) using IEEE 802.11g with 54 Mbps[1]. The access point is connected to the receiver using fast Ethernet with 100 Mbps. The distance between the wireless stations and the access point is between 0.5 m and 1.5 m to avoid antenna near field effects. We switched off RTS/CTS, automatic rate adaption and made sure, that packet fragmentation does not occur. Additionally, all stations are connected to a separated wired control network.

We perform experiments with $M = 2, 3, 4$ contending stations lasting for 30, 45, and 60 minutes respectively. We repeat each experiment 25 times to generate a sufficient amount of data for statistical analysis. Each station sends a greedy UDP flow of 1500 Byte packets at a rate of 28 Mbps. Note that the sending rate of a single station coincides with the net service rate of IEEE 802.11g due to per-packet protocol overhead [2]. During the experiments all stations transmitted their data with an average rate of 14, 9.3, and 7 Mbps respectively indicating close to perfect long-term fairness (2) in the testbed.

To investigate fairness at different timescales we tag a station and count the random number of packet inter-transmissions $K$ of all contending stations while the tagged station transmits $l$ packets, i.e. we consider the conditional probability $\mathsf{P}[K = k|l]$. For $M = 2$ and $l = 1$ our definition reduces to the special case in [4] where short-term fairness among two stations is analyzed. Here, we consider short- as well as long-term fairness and an arbitrary number of stations.

Jain's fairness index (1) derives immediately from the first and second moment of the inter-transmissions. We compute the empirical distribution for different $l$ to address short- and long-term fairness in our testbed. Fig. 2 shows the results for $M = 2, 3, 4$ contending stations and $l = 1 \ldots 100$. It becomes apparent that Jain's index reports poor short-term fairness, i.e. for small $l$ stations frequently outperform each other. This short-term unfairness is, however, quickly alleviated, e.g. we see a fairness index of 0.9 already for $l = 20$ in case of two stations, converging to long-term fairness where all stations virtually sent an equal number of packets. Fairness decreases with increasing number of stations as also measured in [24], possibly due to a subset of stations in exponential backoff [27].

---

[1] We use Lenovo ThinkPad R61 and T61 notebooks with 2.0 GHz, 2 GB RAM running Ubuntu Linux 7.10 with kernel version 2.6.25. We employ the internal Intel PRO/Wireless 4965 AG IEEE 802.11g WLAN cards. The access point is a Buffalo Wireless-G 125 series running DD-WRT [38] version 24.

Next, we compare our testbed measurements to OmNet++ simulations. The simulator assumes perfect channel conditions without interference or fading as also expected in our measurement room. Fig. 3(a) shows the probability mass function (pmf) of the inter-transmissions $K$ for different $l$ to compare the simulations to the testbed measurements. We observe that the simulation results match the measurement data quite well. We use quantile-quantile (q-q) plots to bring out differences at the tail of the distributions to detail the goodness of fit. The exemplary q-q plot for $l = 40$ in Fig. 3(b) shows the quantiles of the measurement data vs. the simulation data. As indicated in Fig. 3(b) we find that 0.99 of the samples coincide, whereas the deviations at the tail show additional unfairness in the testbed that is not reproduced by the simulator.

Furthermore, we compare our testbed measurements to the model that is established for two stations and short-term fairness in [4] and introduced in Sect. II-B. To analyze long-term fairness as well we extend this method from the special case $l = 1$ in [4] to cover all $l \geq 1$. Denote $b_i(j)$ the i.i.d. countdown values for packet $j$ at station $i = 1, 2$. We model the distribution of inter-transmissions $K$ of station 1 while station 2 transmits $l$ packets as

$$\mathsf{P}[K=k|l] = \mathsf{P}\left[\sum_{j=1}^{k} b_1(j) \leq \sum_{j=1}^{l} b_2(j) \text{ and } \sum_{j=1}^{k+1} b_1(j) > \sum_{j=1}^{l} b_2(j)\right].$$

Assuming $b_i(j)$ are uniform random variables as in [4] results in Irwin-Hall distributed terms which, however, do not yield a simple solution for $l \geq 2$. For now, we compute the distribution numerically by convolution and compare the results denoted uniform backoff model in Fig. 4 to our testbed data.

The pmf of the model shows a clear deviation from the testbed results that is also confirmed by the q-q plot. Compared to Fig. 3(b) the 0.99 interval becomes much narrower in Fig. 4(b) indicating that the assumption of uniform backoff values overestimates the actual fairness of the DCF significantly.

## IV. Stochastic Models for Fairness Analysis

In this section we use probability theory, see e.g. [39], to derive closed-form expressions for the fairness that is achieved among $M$ contending stations. We tag station $M$ and denote $K_i$ the inter-transmissions of station $i = 1 \ldots M - 1$ and let $K = \sum_{i=1}^{M-1} K_i$. The conditional probability $\mathsf{P}[K = k | l]$ can be defined for $M \geq 2$ as

$$\mathsf{P}[K = k | l] = \mathsf{P}\left[\sum_{i=1}^{M-1} K_i = k\right] \quad (3)$$

where the random variables $K_i$ are the integers that satisfy

$$\sum_{j=1}^{K_i} b_i(j) \leq \sum_{j=1}^{l} b_M(j) \quad \text{and} \quad \sum_{j=1}^{K_i+1} b_i(j) > \sum_{j=1}^{l} b_M(j).$$

First, in Sect. IV-A we analyze long-term fairness among two stations, i.e. $M = 2$ and $l, k \gg 1$. Assuming exponential backoff we derive results for arbitrary $M, l, k$ in Sect. IV-B.

### A. Long-term Fairness with Arbitrary Backoff

We use the central limit theorem to derive the long-term fairness. In the sequel, we denote normal random variables $N(\mu, \sigma^2)$ where $\mu$ is the mean and $\sigma^2$ the variance.

**Theorem 1 (Gaussian approximation)** *Let the $b_i(j)$ be i.i.d. random variables with mean $\mu$ and variance $\sigma^2$ and let $M = 2$. For $k, l \gg 1$ (3) is approximately Gaussian where*

$$\mathsf{P}[K \leq k | l] \approx \mathsf{P}\left[N(0,1) \leq \frac{\mu(k-l)}{\sigma\sqrt{k+l}}\right].$$

*Proof:* For $M = 2$ we have from (3) that

$$\mathsf{P}[K < k | l] = \mathsf{P}\left[\sum_{j=1}^{k} b_1(j) > \sum_{j=1}^{l} b_2(j)\right]$$

and after expansion and some normalization this equals

$$= \mathsf{P}\left[\frac{\sum_{j=1}^{l} b_2(j) - l\mu}{\sigma\sqrt{l}} - \frac{\sum_{j=1}^{k} b_1(j) - k\mu}{\sigma\sqrt{l}} < \frac{\mu(k-l)}{\sigma\sqrt{l}}\right].$$

Using the central limit theorem it follows that

$$\mathsf{P}[K < k | l] \approx \mathsf{P}\left[N(0,1) - N\left(0, \frac{k}{l}\right) < \frac{\mu(k-l)}{\sigma\sqrt{l}}\right].$$

Since the normal distribution with zero mean is symmetric we can replace the subtraction of $N(0, k/l)$ by addition. Furthermore, the sum of two normal random variables $N(\mu_1, \sigma_1^2)$ and $N(\mu_2, \sigma_2^2)$ is normal with $N(\mu_1 + \mu_2, \sigma_1^2 + \sigma_2^2)$ such that

$$\mathsf{P}[K < k | l] \approx \mathsf{P}\left[N\left(0, \frac{k+l}{l}\right) < \frac{\mu(k-l)}{\sigma\sqrt{l}}\right].$$

Finally, we use that if $X$ is $N(a\mu, a^2\sigma^2)$ then $Y = X/a$ is $N(\mu, \sigma^2)$ with $a^2 = (k+l)/l$ to standardize the result. ∎

Th. 1 assumes i.i.d. random countdown values. It does, however, not make any assumption about their distribution. To compare the impact of different models we formulate the following corollary for uniform, as used e.g. in the short-term fairness model in [4], respectively exponential countdown values, as assumed e.g. in the throughput model in [2], [31].

**Corollary 1 (Uniform versus exponential countdown)**
*Assume Th. 1. If the $b_i(j)$ are uniform in $[0, w]$, then*

$$\mathsf{P}[K \leq k | l] \approx \mathsf{P}\left[N(0,1) \leq \frac{\sqrt{3}(k-l)}{\sqrt{k+l}}\right].$$

*If the $b_i(j)$ are exponentially distributed then*

$$\mathsf{P}[K \leq k | l] \approx \mathsf{P}\left[N(0,1) \leq \frac{k-l}{\sqrt{k+l}}\right].$$

Th. 1 and Cor. 1 yield a number of important conclusions. First, we compare the pmf from Cor. 1 displayed in Fig. 5(a). The assumption of exponential backoff values, which reflects the increase of the contention window in case of collisions, matches our empirical data closely. In contrast the assumption of uniform backoff does not agree well. Fig. 5(c) and Fig. 5(d) show q-q plots of the exponential model vs. the testbed data and confirm the accuracy of the model for 0.99 of the samples. The testbed exhibits larger unfairness at the distribution tail as also observed compared to OmNet++ simulations in Fig. 3(b).

Considering Cor. 1 it is interesting to note that the distribution parameter in case of uniform as well as in case of exponential countdown values has no influence on the fairness. In contrast the distribution itself has significant impact. Cor. 1 shows an explicit fairness degradation of $\sqrt{3}$ of exponential compared to uniform countdown values, i.e. $\sqrt{3}$ can be viewed as the price of exponential backoff. Fig. 5(a) shows this effect clearly, i.e. the pmf for uniform is by $\sqrt{3}$ higher and narrower.

Another important aspect is the improvement of long- over short-term fairness. To this end, we define a multiplicative constant $c$ and let $k = cl$. The parameter $c$ may be viewed as a threshold value that specifies a relative deviation that is still considered fair. By insertion the term $(k - l)/\sqrt{k + l}$ from Th. 1 becomes $\sqrt{l}(c - 1)/\sqrt{c + 1}$ and it follows that long-term fairness improves proportionally to $\sqrt{l}$. Thus, the initial fairness improvement for small $l$ is significant but becomes less pronounced with increasing $l$. This result is independent of the distribution of backoff values. Fig. 5(b) shows a q-q plot of our measurement data for $l = 40$ vs. $l = 160$. The slope of the q-q plot closely follows $\sqrt{160/40}$ for 0.99 of the samples. Hence, Fig. 5(b) clearly displays the $\sqrt{l}$ scaling in the testbed data.

### B. Short- and Long-term Fairness with Exponential Backoff

In the sequel we only consider exponential backoff that proved accurate in Sect. IV-A. We derive an exact result and useful approximations for long- as well as short-term fairness.

**Theorem 2 (Exact result)** *Let $b_i(j)$ be i.i.d. exponential random variables and let $p = 1/M$. Then (3) is negative binomial*

$$\mathsf{P}[K = k | l] = p^l (1-p)^k \binom{k + l - 1}{k}.$$

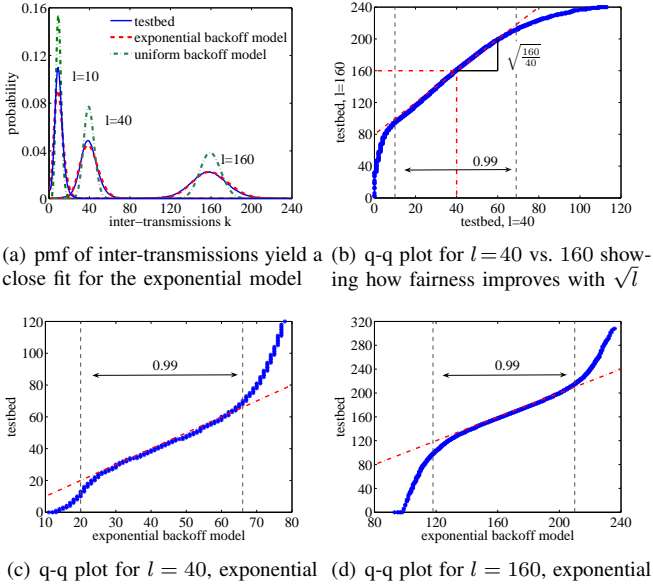

(a) pmf of inter-transmissions yield a close fit for the exponential model

(b) q-q plot for $l=40$ vs. 160 showing how fairness improves with $\sqrt{l}$

(c) q-q plot for $l = 40$, exponential

(d) q-q plot for $l = 160$, exponential

Fig. 5. Analytical results from Cor. 1 match the testbed data well, if the model uses exponential instead of uniform backoff random variables. The model accurately predicts the $\sqrt{l}$ fairness improvement, see the slope in 5(b).

*Proof:* First, we determine the probability that station $M$ gets access to the channel. Key to deriving this probability is the memorylessness of negative exponential random variables, i.e. given an exponential random variable $X$ with parameter $\lambda$ it holds that $\mathsf{P}[X > x + y | X > x] = \mathsf{P}[X > y] = e^{-\lambda y}$.

Consider $M$ stations that contend for an idle channel. Owing to the memoryless property each station has an exponentially distributed countdown value with the same parameter $\lambda$ irrespective of the time the station has already spent on performing the countdown procedure. It follows that each channel access can be viewed as an independent Bernoulli experiment. Denote $p$ the probability of success, i.e. the probability that station $M$ finishes its countdown procedure first such that it attains access to the channel. Since the remaining countdown values are i.i.d. at all stations, each station has the same channel access probability $p$ where $\sum_{i=1}^{M} p = 1$ such that $p = 1/M$.

Concluding, the probability that stations $1 \ldots M-1$ access the channel exactly $k$ times until station $M$ performs the $l$-th channel access is the probability to see the $l$-th success of station $M$ exactly in the $(k+l)$-th Bernoulli trial. This event is negative binomially distributed. ■

A direct proof from (3) for $M = 2$ can be found in the appendix. Next, we derive a useful bound for the distribution.

**Corollary 2 (Chernoff bound)** *Assume Th. 2. It follows that*

$$\mathsf{P}\left[K \lesseqgtr k \Big| l\right] \leq \left(\frac{(1-p)(k+l)}{k}\right)^k \left(\frac{p(k+l)}{l}\right)^l \quad \forall k \lesseqgtr l(M-1)$$

*Proof:* The proof uses Chernoff's bounds

$$\mathsf{P}\left[X \lesseqgtr x\right] \leq e^{-\theta x} \mathsf{M}_X(\theta) \quad \forall \theta \lesseqgtr 0$$

where $\mathsf{M}_X(\theta) = \mathsf{E}[e^{\theta X}]$ denotes the moment generating function of $X$. We insert the well-known generating function of the negative binomial random variable and derive

$$\mathsf{P}[K \leq k | l] \leq e^{-\theta k} \left(\frac{p}{1-(1-p)e^{\theta}}\right)^l \quad \forall \theta < 0.$$

We minimize the right hand side over all $\theta < 0$ to obtain the best possible bound and insert $\theta = \ln k - \ln((l+k)(1-p))$ where $k < l(1-p)/p = l(M-1)$ to ensure $\theta < 0$. The upper bound follows in the same way. ■

Finally, we derive a normal approximation. This allows viewing fairness deviations as i.i.d. Gaussian noise.

**Corollary 3 (Gaussian approximation)** *Assume Th. 2. It follows for $l \gg 1$ that*

$$\mathsf{P}[K \leq k | l] \approx \mathsf{P}\left[N(0,1) \leq \frac{kp - l(1-p)}{\sqrt{l(1-p)}}\right].$$

*Proof:* We view the negative binomial random variable in Th. 2 as a sum of i.i.d. geometric random variables. Each of the geometric random variables equals the number of trials required until the next success is achieved, i.e. we write Th. 2 as a sum of $l$ i.i.d. geometric random variables denoted $X_i$

$$\mathsf{P}[K \leq k | l] = \mathsf{P}\left[\sum_{i=1}^{l} X_i \leq k + l\right].$$

Normalization using the mean $\mu = 1/p$ and the variance $\sigma^2 = (1-p)/p^2$ of geometric random variables yields

$$\mathsf{P}[K \leq k | l] = \mathsf{P}\left[\frac{\sum_{i=1}^{l} X_i - l\mu}{\sigma\sqrt{l}} \leq \frac{(k+l)p - l}{\sqrt{l(1-p)}}\right].$$

Using the central limit theorem the normalized sum is approximately standard normal $N(0,1)$ if $l \gg 1$. ■

We note that Cor. 1 and Cor. 3 for $M = 2$ converge under the assumption of the central limit theorem, i.e. for $k, l \to \infty$.

Th. 2, Cor. 2, and Cor. 3 can be easily extended to heterogenous stations that use different parameters $\lambda$, e.g. for service differentiation. In this case only the probability of successful channel access $p$ has to be adapted accordingly. Also, it is a straightforward extension of Th. 2 to derive the probability that a single station with index $M$ transmits $l$ packets given that the remaining $M-1$ stations together transmit $k$ packets.

Fig. 6 compares the results from Th. 2 with the testbed measurement data. We find that Th. 2 accurately predicts short- and long-term fairness for $M = 2$ stations, see Fig. 6(a). Fig. 6(c) and Fig. 6(d) add q-q plots for short-term fairness which also show a close match for 0.99 of the samples. For three respectively four stations we find that Th. 2 underestimates unfairness, see Fig. 6(b). While our model fits almost perfectly for two stations we conclude that additional effects beyond our model cause unfairness in case of more than two stations. Empirical results indicating poor fairness in case of more than two stations have also been reported e.g. in [24].

Considering Th. 2 we recover the result that the parameter of the i.i.d. exponentially distributed countdown values does not impact fairness regardless of the number of stations and the time-scale.

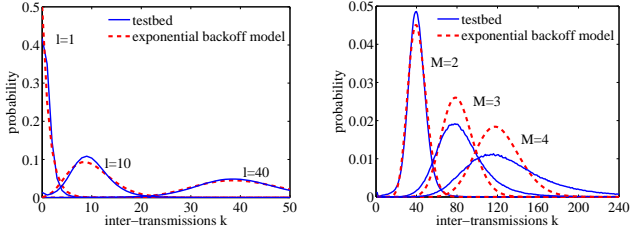

(a) pmf of inter-transmissions, $M=2$ (b) pmf of inter-transmissions, $l=40$

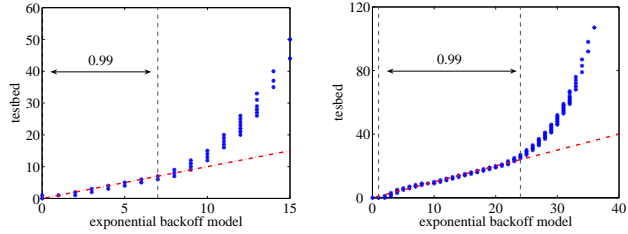

(c) q-q plot for $l=1$ and $M=2$ (d) q-q plot for $l=10$ and $M=2$

Fig. 6. Analytical results from the exponential model and Th. 2 predict short- and long-term fairness correctly for $M=2$ stations. For more stations the testbed data shows additional unfairness beyond the model, see 6(b).

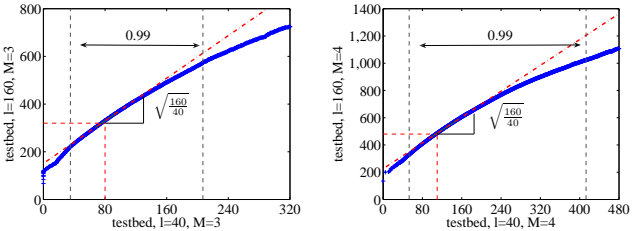

(a) q-q plot for $l=40$ vs. $160$, $M=3$ (b) q-q plot for $l=40$ vs. $160$, $M=4$

Fig. 7. Long-term fairness improves approximately with $\sqrt{l}$ in case of more than two stations. This improvement is correctly predicted by Cor. 3.

With $p=1/M$ and letting $k=(M-1)cl$ Cor. 3 yields

$$\frac{kp - l(1-p)}{\sqrt{l(1-p)}} = \sqrt{l}(c-1)\sqrt{\frac{M-1}{M}}$$

predicting that fairness improves with $\sqrt{l}$. Fig. 7(a) and 7(b) show the improvement for $M=3$ and $M=4$ stations respectively. The testbed measurements confirm the dependence on $\sqrt{l}$. Fairness even scales better than $\sqrt{l}$ at the tail end.

Next, we derive Jain's fairness index (1) that follows directly from the first and second moment of the inter-transmissions. For the negative binomial distribution in Th. 2 the first moment is $\mathsf{E}[K] = l(1-p)/p$ and the second central moment is $l(1-p)/p^2$ such that the second moment becomes $\mathsf{E}[K^2] = l(1-p)/p^2 + (l(1-p)/p)^2$. With $p=1/M$ we have

$$f = \frac{l}{l + \frac{M}{M-1}}. \quad (4)$$

Regarding Fig. 2 we find that (4) matches the testbed measurement data almost perfectly for $M=2$ stations, in which case $f = l/(l+2)$. For more than two stations additional effects that cause unfairness as shown in Fig. 6(b) result in a deviation. In face of short-term unfairness we find, however, that fairness is practically achieved already for moderate $l$.

For further quantitative comparison we use the Kullback-Leibler distance between the measurement data and the analytical expressions. The Kullback-Leibler distance $\mathsf{D}(X||Y)$ quantifies the deficiency if we assume the distribution of $X$ instead of the true distribution of $Y$ [40]. It is defined as

$$\mathsf{D}(X||Y) = \sum_x \mathsf{P}(X=x) \ln \frac{\mathsf{P}(X=x)}{\mathsf{P}(Y=x)}.$$

Here, $\mathsf{P}(X=x)$ is the empirical mass function and $\mathsf{P}(Y=y)$ is the probability mass function from the analytical model. Tab. I summarizes the results. Clearly, modeling the backoff procedure using exponential random variables outperforms the assumption of uniform random variables. Moreover, the results confirm that fairness degrades in case of more than 2 stations.

## TABLE I
KULLBACK-LEIBLER DISTANCE FOR DIFFERENT MODELS

| model | | $l=1$ | $l=10$ | $l=40$ | $l=160$ |
|---|---|---|---|---|---|
| uniform backoff | Cor. 1 | NA | 0.536 | 0.766 | 0.831 |
| exponential backoff | Cor. 1 | NA | 0.104 | 0.094 | 0.075 |
| exponential backoff | Th. 2 | 0.039 | 0.050 | 0.083 | 0.065 |

| model | | $M=2$ | $M=3$ | $M=4$ |
|---|---|---|---|---|
| exponential backoff | Th. 2 | 0.083 | 0.351 | 0.702 |

## V. DCF CLOCK AND A SERVICE CURVE MODEL

In this section we derive a service curve model for the DCF that facilitates applications of the stochastic network calculus [18], [19], [20]. To this end, we model the DCF as emulating the Generalized Processor Sharing (GPS) discipline [7].

### A. The DCF viewed as a GPS Emulation

GPS is a fluid-flow model that defines a weighted fair resource allocation. Each flow indexed $i$ is assigned a weight $\varphi_i$. Considering only backlogged flows, flow $i$ is guaranteed a share of $\varphi_i / \sum_j \varphi_j$ of the capacity $C$. Due to the granularity of packets real implementations can only emulate GPS with limited precision, for an overview see e.g. [11], [12], [13] and for enhancements of the DCF [8], [9], [10]. Analytical models specify the deviation from an ideal GPS system using worst-case error terms. Two prominent models are Guaranteed Rate Clock (GRC) [11] and Packet Scale Rate Guarantee (PSRG) [14] that are the basis of IntServ and DiffServ. We derive a related model for the DCF that we refer to as DCF Clock.

**Lemma 1 (DCF Clock)** *Consider $M$ stations with indices $i$ that contend for the medium using the DCF. Let $a_i(n)$ be the arrival time of the $n$-th packet at station $i$ with length $L_i$. If the medium is busy at $a_i(n)$ let $\delta$ be the residual service time of the packet in service or else $\delta = 0$. The departure times are*

$$d_i(n) = \max\{a_i(n) + \delta, d_i(n-1)\} + \frac{L_i}{R_i} + \phi_i(n) + \psi_i(n)$$

*where the average service rate*

$$R_i = \frac{L_i}{C(\mu + M\Delta) + \sum_{j=1}^{M} L_j} C$$

*is subject to zero mean error terms* $\phi_i(n) = b_i(n) - \mu$ *and*

$$\psi_i(n) = \sum_{j=1}^{M} (K_j(n) - 1)\left(\frac{L_j}{C} + \Delta\right).$$

*Here, $C$ is the capacity, $b_i(n)$ are i.i.d. exponential countdown values with mean $\mu$, $K_j(n)$ are the inter-transmissions, and $\Delta$ comprises all constant per-packet protocol latencies.*

*Proof:* First assume that station $i$ is not backlogged. If the medium is idle station $i$ starts the medium access procedure immediately at $a_i(n)$ or else after the residual service time of the packet that is in service, i.e. at $a_i(n) + \delta$. Otherwise, if station $i$ is backlogged it starts the access procedure for packet $n$ after packet $n-1$ finishes service, that is at $d_i(n-1)$. Combining all cases station $i$ initiates its access procedure at $\max\{a_i(n) + \delta, d_i(n-1)\}$.

Before station $i$ transmits packet $n$ an amount of channel idle time of $b_i(n)$ has to be accumulated to complete the countdown procedure. In parallel all other backlogged stations perform their countdown procedure to contend for the medium.

The transmission of a packet under the DCF includes constant protocol overhead for DIFS, preamble, SIFS, and acknowledgement that are summed up in $\Delta$. Hence, it takes $\Delta + L_i/C$ units of time to transmit a packet of length $L_i$ on a channel with capacity $C$. The number of packets transmitted by station $j$ in the interval $[\max\{a_i(n) + \delta, d_i(n-1)\}, d_i(n)]$ is denoted $K_j(n)$. The transfer takes $\sum_{j=1}^{M} K_j(n)(\Delta + L_j/C)$ units of time. Assembling all parts the departure time is

$$d_i(n) = \max\{a_i(n) + \delta, d_i(n-1)\} + b_i(n) + \sum_{j=1}^{M} K_j(n)\left(\frac{L_j}{C} + \Delta\right). \quad (5)$$

Next, we show that the two error terms have zero mean. Clearly, $\mathsf{E}[\phi_i(n)] = \mathsf{E}[b_i(n) - \mu] = 0$. We instantiate Th. 2 with $M = 2$ and $l = 1$ to find the number of inter-transmissions of one station. We have $\mathsf{E}[K_j(n)] = 1$ such that $\mathsf{E}[\psi_i(n)] = 0$.

The residuum after substitution of the error terms in (5) is the mean latency caused by the countdown procedure and by inter-transmissions. We equate this latency with $L_i/R_i$ where $R_i$ has the interpretation of an average service rate. We obtain

$$\mu + \sum_{j=1}^{M}\left(\frac{L_j}{C} + \Delta\right) = \frac{L_i}{R_i}$$

and solve for $R_i$ to derive the average service rate. ∎

Lemma 1 specifies the deviation $\phi_i + \psi_i$ of the DCF clock from an ideal GPS system with rate allocation $R_i$. Note that the error terms have zero mean such that $R_i$ is the true average service rate. The service rate considers the resource consumption that is due to protocol overhead. Apart from that, the rate allocation is proportional to the packet lengths used by individual stations, which formally derives from the target of packet-level fairness of the DCF. Roughly speaking, the packet lengths $L_i$ take the place of the GPS weights $\varphi_i$.

It is worthwhile comparing the error terms in Lemma 1 with the GRC model [11]. The GR Clock is defined as

$$GRC_i(n) = \max\{a_i(n), GRC_i(n-1)\} + \frac{L_i}{R_i}$$

where departures are subject to an error term $\chi$ such that

$$d_i(n) \leq GRC_i(n) + \chi.$$

Compared to the GR Clock the recursion in Lemma 1 uses the actual departure times $d_i$ instead of the target $GRC_i$. As a consequence, the per-packet error terms of the DCF are accumulated during a busy period. This is not the case in the GRC model. In other words, a GRC scheduler that deviates from the GR Clock nevertheless has to keep up with the GR Clock at subsequent packet transmissions, i.e. unfairness cannot accumulate. In contrast the DCF does not seek to correct previous deviations, i.e. the DCF is memoryless in the sense that it does not compensate past unfairness. Moreover, the error terms of known GRC implementations typically are small deterministic upper bounds, e.g. $\chi = L_{\max}/C$ for packet-by-packet GPS [11] as opposed to the DCF error terms that are random and possibly unbounded.

### B. A Stochastic Service Curve Model of the DCF

Service curves are a powerful model for systems in the network calculus, see the textbooks [16], [17] for an overview. The particular strength is the convolution of tandem systems that yields the notion of network service curve and permits analyzing entire networks as a single system. Recent stochastic network calculus is developed in [18], [19], [20] and references therein. The stochastic network calculus facilitates analyzing wireless systems. It is used in [21], [20], [22] to model the effects that are due to fading and interference.

In this section we derive a stochastic service curve model for the DCF. We use a max-plus approach, see e.g. [16], [41], that translates to min-plus network calculus if the service curve is inverted from a function of packets into a function of time [16]. In the deterministic network calculus this relation establishes a close connection between the GRC model and so-called latency-rate service curves [12], [13], [42].

**Definition 1 (Stochastic max-plus service curve)** Consider a system with packet arrival and departure times $a(n)$ and $d(n)$ respectively. The system has a stochastic max-plus service curve $s_\varepsilon$ with violation probability $\varepsilon$ if for all $n \geq 1$

$$\mathsf{P}[d(n) \leq a * s_\varepsilon(n)] \geq 1 - \varepsilon.$$

Here, $*$ denotes the max-plus convolution that is defined as $a * s_\varepsilon(n) = \max_{m \in [1,n]}\{a(n-m+1) + s_\varepsilon(m)\}$.

Given a stochastic max-plus service curve it is straightforward to compute packet delays defined as $d(n) - a(n)$ that are violated at most with probability $\varepsilon$ from $a * s_\varepsilon(n) - a(n)$, e.g. to determine the playout delay of a video application.

**Theorem 3 (DCF service curve)** *Assume Lemma 1, let all packets have the same size denoted L, and consider a tagged flow. The DCF has a stochastic latency-rate service curve*

$$s_\varepsilon(n) = T + \frac{n}{R}$$

*with latency T and rate R defined as*

$$T = \tau + (1+\varsigma)\left(\frac{L}{C} + \Delta\right) \text{ and } \frac{1}{R} = \vartheta + (1+\rho)\left(\frac{L}{C} + \Delta\right)$$

*and violation probability* $\varepsilon = \sum_{m=1}^\infty (\varepsilon_1(m) + \varepsilon_2(m))$. *Parameters* $\tau, \vartheta, \varsigma, \rho \geq 0$ *and* $\varepsilon_1, \varepsilon_2$ *are defined in (7) and (8).*

The service curve in Th. 3 is an affine function that comprises a latency offset $T$ and a packet rate $R$ respectively per-packet latency $R^{-1}$. The terms correspond to the latency-rate service curve model in min-plus algebra. The free parameters define the service guarantee and determine its violation probability. The parameter choice is subject to numerical optimization. We find that $\tau, \vartheta$ that stem from the variable duration of the countdown procedure have comparably small impact, whereas $\varsigma, \rho$ that consider the random amount of time consumed by inter-transmissions have a significant effect.

*Proof:* We consider station $M$ and denote its arrivals $a(n)$ and departures $d(n)$. We analyze a single busy period starting at $a(m)$. From (5) we obtain by recursion that

$$d(n) = a(m) + \delta + \sum_{i=m}^n b(i) + (l+K)\left(\frac{L}{C} + \Delta\right) \quad (6)$$

where $l = n - m + 1$ is the number of packets sent by station $M$ and $K = \sum_{j=1}^{M-1}\sum_{i=m}^n K_j(i)$ is the sum of all inter-transmissions since the start of the busy period.

We derive probabilistic affine upper envelopes for the random terms. $\sum_{i=m}^n b(i)$ is the sum of $l$ i.i.d. exponential random variables each with mean $\mu$. Hence, the sum is Gamma distributed and has moment generating function $\mathsf{M}_{\sum b}(\theta, l) = (1/(1-\theta\mu))^l$ for $\theta < 1/\mu$. From Chernoff's bound we obtain

$$\mathsf{P}\left[\sum_{i=m}^n b(i) \geq \tau + \vartheta l\right] \leq e^{-\theta(\tau+\vartheta l)}\left(\frac{1}{1-\theta\mu}\right)^l \forall \theta \in (0, 1/\mu).$$

Minimization yields $\theta = (\tau + (\vartheta - \mu)l)/((\tau + \vartheta l)\mu)$ such that $\mathsf{P}[\sum_{i=m}^n b(i) \geq \tau + \vartheta l] \leq \varepsilon_1(l)$ where

$$\varepsilon_1(l) = \left(\frac{\vartheta + \frac{\tau}{l}}{\mu} e^{-\frac{\mu+\vartheta+\frac{\tau}{l}}{\mu}}\right)^l. \quad (7)$$

We use Cor. 2 to bound the number of inter-transmissions $K$. We let $k = \varsigma + \rho l$ to find $\mathsf{P}[K \geq \varsigma + \rho l] \leq \varepsilon_2(l)$ where

$$\varepsilon_2(l) = \left(\frac{p(1-p)^{\rho+\frac{\varsigma}{l}}(1+\rho+\frac{\varsigma}{l})^{1+\rho+\frac{\varsigma}{l}}}{(\rho+\frac{\varsigma}{l})^{\rho+\frac{\varsigma}{l}}}\right)^l. \quad (8)$$

Inserting the envelopes $\tau + \vartheta l$ and $\varsigma + \rho l$ into (6) and bounding $\delta$ by $\Delta + L/C$ yields that

$$d(n) \leq a(m) + \tau + \vartheta l + (1 + \varsigma + (1+\rho)l)\left(\frac{L}{C} + \Delta\right) \quad (9)$$

is violated at most with probability $\varepsilon_1(l) + \varepsilon_2(l)$.

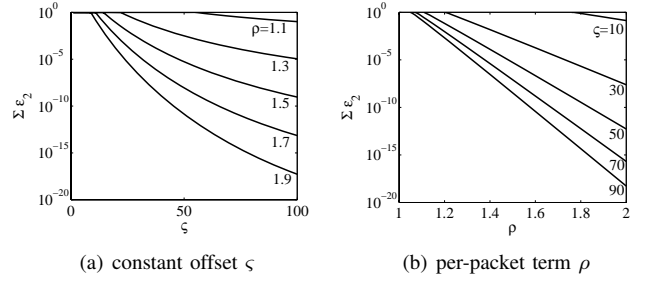

(a) constant offset $\varsigma$   (b) per-packet term $\rho$

Fig. 8. The free service curve parameters $\varsigma, \rho$ define a sample path bound for the number of inter-transmissions. The bound is violated at most with probability $\sum \varepsilon_2$. The parameters have significant impact on the shape of the latency-rate service curve.

In the final step we take the maximum over all $m$ of the right hand side in (9) to resolve the assumption that the busy period starts at $a(m)$ [16], [17] and obtain the max-plus convolution form in Def. 1. After some reordering (9) yields the service curve in Th. 3. We use Boole's inequality and sum $\varepsilon_i(l)$ over all $l$ to derive a corresponding sample path bound [18]. We estimate the tail probabilities to verify that this sample path bound exists. From (7) we have

$$\varepsilon_1(l) \leq \left(\frac{\vartheta}{\mu} e^{-\frac{\mu+\vartheta}{\mu}}\right)^l = q_1^l$$

where $q_1 < 1$ generally. Similarly, we have from (8) that

$$\varepsilon_2(l) \leq \left(\frac{p(1-p)^\rho (1+\rho)^{1+\rho}}{\rho^\rho}\right)^l = q_2^l$$

where $q_2 < 1$ can be shown for $\rho > (1-p)/p = M - 1$ from Bernoulli's inequality. Using the geometric sum we find that

$$\sum_{l=n}^\infty \varepsilon_i(l) \leq \frac{q_i^n}{1 - q_i}$$

proving the boundedness of the violation probability. ■

We provide an example of the service curve for IEEE 802.11g. We set $\Delta = 0.1$ ms and $C = 54$ Mbps. We consider $M = 2$ stations and packets of $L = 1500$ Byte. The parameters $\tau, \vartheta$ have comparably small impact and $\tau = 1$ ms and $\vartheta = 0.1$ ms achieve already $\sum \varepsilon_1 < 10^{-6}$. The effects of the remaining parameters $\varsigma, \rho$ on $\sum \varepsilon_2$ are shown in Fig. 8. As an example consider $\varsigma = 50$ and $\rho = 1.5$ where Fig. 8 reveals $\sum \varepsilon_2 \approx 5 \cdot 10^{-6}$. The corresponding service curve is $s_\varepsilon(n) = 17.5 + 0.9n$ ms where $\varepsilon = \sum \varepsilon_1 + \sum \varepsilon_2 \approx 6 \cdot 10^{-6}$.

## VI. Rate Estimation from Passive Measurements

In this section we show how user applications, such as rate-adaptive video streaming, can estimate their fair bandwidth share under the DCF from passive measurements of their data arrivals and departures. Closely related are active probing techniques that seek to identify the unused capacity along a network path from specific probe packets, also referred to as available bandwidth estimation. For an overview of these methods and their use in IEEE 802.11 WLANs we refer to [35] and references therein. Compared to available bandwidth estimation two major differences are important to consider when estimating the fair rate under the DCF.

The majority of the available bandwidth estimation methods are based on the assumption of First-In First-Out (FIFO) scheduling. Due to its simplicity FIFO is the prevalent scheduling discipline in the wired Internet. Yet, packet probes can preempt existing production traffic at a FIFO multiplexer making the estimation of unused capacity challenging. This difficulty does not occur in case of fair scheduling where a greedy flow is simply assigned the fair bandwidth share.

For the moment assume an ideal GPS system and a flow that transmits a burst of $l+1$ packets. Packets are marked with time-stamps $a(n)$ and $d(n)$ at the sender and receiver respectively. To avoid the necessity for synchronized clocks at sender and receiver we use only the time differences between packet departures, so-called gaps, that can be computed from the time-stamps at the receiver only. The averaged gap of the departures is

$$g_d = \frac{d(l+1) - d(1)}{l}$$

and an estimate of the fair rate follows as $f = L/g_d$.

In case of the DCF such fair rate estimates can, however, be largely perturbed by the random channel access procedure. We view this variability as measurement noise and use a Kalman filter to perform an online smoothing over several samples of $g_d$. The Kalman filter takes measurements of the departures of a system superposed by i.i.d. Gaussian noise as input to estimate the system's state. It recursively weights past measurements to generate optimal estimates in the sense that it minimizes the mean squared error. Stationarity of the system is not a necessary precondition. For an overview see e.g. [43].

We use a Kalman filter to generate smoothed estimates of the departure gap. We index consecutive measurement samples $g_d(n)$ and denote the smoothed output of the Kalman filter $\bar{g}_d(n)$. The Kalman filter computes

$$\bar{g}_d(n) = (1 - G(n))\bar{g}_d(n-1) + G(n)g_d(n)$$

where the weight $G(n)$ balances the impact of the current measurement. $G(n)$ is referred to as the Kalman gain that is

$$G(n) = \left(1 + \frac{\sigma_{g_d}^2(n)}{\sigma_E^2(n-1) + \sigma_P^2}\right)^{-1}$$

where $\sigma_E^2$ is the estimation error variance and $\sigma_P^2$ and $\sigma_{g_d}^2$ are external parameters denoting Gaussian process and measurement noise variances respectively. The update of $\sigma_E^2$ is

$$\sigma_E^2(n) = (1 - G(n))(\sigma_E^2(n-1) + \sigma_P^2).$$

To parameterize the measurement noise, we estimate the variability of $g_d$. From (6) we derive that

$$d(l+1) - d(1) = (l+K)\left(\frac{L}{C} + \Delta\right) + \sum_{j=2}^{l+1} b(j) \quad (10)$$

where $K$ are the inter-transmissions between packet 1 and $l+1$. Eq. (10) contains two sources of randomness, the number of inter-transmissions from contending stations $K$ and the cumulated countdown values $\sum_{j=2}^{l+1} b(j)$. As discussed in Sect. V-B the effects due to inter-transmissions are dominant so that we approximate $\sum_{j=2}^{l+1} b(j) \approx l\mu$ by its mean value. Regarding the inter-transmissions we assume $l \gg 1$ and use the Gaussian model from Cor. 3. After normalization we find that $K/l$ is normal with standard deviation $\sigma_{K/l} = \sqrt{(1-p)/(p^2 l)}$ where $p = 1/M$ is determined by the number of contending stations. Dividing (10) by $l$ we recover $g_d$ and derive the standard deviation of the gap as $\sigma_{g_d} = \sqrt{(1-p)/(p^2 l)} \, (\Delta + L/C)$.

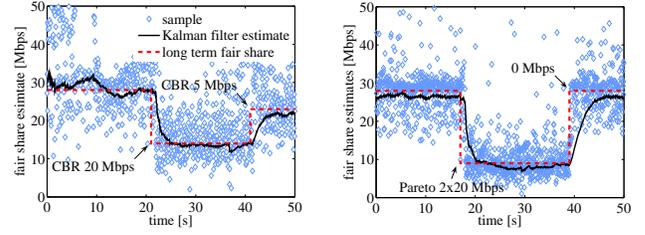

(a) video with CBR cross traffic  (b) video with Pareto cross traffic

Fig. 9. Samples of the fair share show a large variability referred to as measurement noise. The noise reduces with $\sqrt{l}$. With this input the Kalman filter provides good estimates of the fair share after short convergence both for CBR as well as Pareto cross traffic and for several contending stations.

To demonstrate the filtering approach we employ rude/crude to transmit a variable bit rate video tracefile[2] using UDP. In addition the D-ITG traffic generator [45] is used at contending stations to generate CBR respectively Pareto cross traffic with a shape parameter of 1.4 and changing intensity for both types. We use each of the packet bursts caused by the video frames to obtain a sample of the fair share. The samples are fed into a Kalman filter that generates smoothed fair share estimates. In our experiments we optimize the filter configuration for $M = 2$ and $\Delta + L/C \approx 0.32$ ms such that the measurement noise has $\sigma_{g_d} \approx 0.46$ ms/$\sqrt{l}$. The Kalman filter weights each of the samples according to the corresponding measurement noise. The adjustment of the process noise as a second free parameter of the Kalman filter can be viewed as a tuning knob that trades smoothness for convergence speed. Here, a process noise with $\sigma_P^2 = 10^{-5}$ ms$^2$ provides good results.

Fig. 9 shows the individual samples derived from each of the video frames as well as smoothed fair share estimates from the Kalman filter. Despite the large variability of single fair share samples the Kalman estimates follow the theoretical long term fair share from (2) closely. Furthermore, the filter quickly detects changes in the underlying process, i.e. changes of the cross traffic that are labeled with traffic type and intensity in Fig. 9. This demonstrates the utility of our approach.

## VII. CONCLUSIONS

We analyzed the short- and long-term fairness of the DCF in IEEE 802.11 based on conditional probabilities of the number of inter-transmissions. The approach has proven highly useful, facilitating significant closed-form results. Regarding i.i.d. countdown values we showed a major impact of the type but not the parameters of the distribution. We proved that long-term fairness improves with $\sqrt{l}$. Our findings are substantiated by an extensive measurement and simulation study. We modeled the DCF as emulating a fluid GPS scheduler yielding a fair average service rate that is subject to well-defined error terms. Based on the DCF clock we derived a service curve that opens up significant options for performance analysis of wireless multi-hop networks using the stochastic network calculus. We concluded our study showing a technique that estimates the fair rate under the DCF from passive traffic measurements of a video application, e.g. for rate-adaptation.

---

[2]We used 140 seconds of a H.264 tracefile from Terminator 2 [44] with an average rate of 5 Mbps and significant variability. The GoP size is 12, the minimum frame size 51 Byte, and the maximum frame size 159 kByte.

### ACKNOWLEDGEMENTS

We would like to thank Z. Bozakov and A. Rizk for many fruitful discussions and M. Hollick, R. Steinmetz, and R. Jakoby for providing the equipment and facilities that made this research possible. This work has been funded by an Emmy Noether grant of the German Research Foundation (DFG).

### APPENDIX

TABLE II

SUMMARY OF NOTATION

| symbol | definition |
|---|---|
| D | Kullback-Leibler distance |
| E | expected value |
| M | moment generating function |
| P | probability |
| $a$ | packet arrival time |
| $b$ | backoff countdown value |
| $c$ | constant fairness threshold |
| $C$ | channel capacity |
| $d$ | packet departure time |
| $E$ | estimation error for the Kalman filter |
| $f$ | fair share, fairness index |
| $g$ | gap, time between packets |
| $G$ | gain of the Kalman filter |
| $GRC$ | Guaranteed Rate Clock |
| $i, j$ | running indices |
| $k$ | number of packet inter-transmissions |
| $K$ | random number of packet inter-transmissions |
| $l$ | number of packets transmitted by a tagged station |
| $L$ | packet length |
| $m, n$ | packet indices |
| $M$ | number of contending stations |
| $N$ | Normal random variable |
| $p$ | channel access probability |
| $P$ | process noise for the Kalman filter |
| $q$ | probability |
| $r$ | source data rate |
| $R$ | data rate of a service curve |
| $s$ | max-plus service curve |
| $T$ | latency of a service curve |
| $w$ | backoff window size |
| $X, Y, Z$ | random variables |
| $\delta$ | residual packet service time |
| $\Delta$ | constant per-packet protocol overhead |
| $\varepsilon$ | violation probability |
| $\theta$ | free parameter in Chernoff's bound |
| $\lambda$ | parameter of the exponential distribution |
| $\mu, \sigma$ | mean and standard deviation |
| $\tau, \vartheta, \varsigma, \rho$ | service curve parameters |
| $\phi, \psi$ | DCF Clock error terms |
| $\chi$ | GR Clock error term |
| $\varphi$ | GPS weights |

### A. Proof of Theorem 2

*Proof:* We provide an alternative proof of Theorem 2 which we derive directly from (3) for $M = 2$. In this case we have

$$\mathsf{P}[K=k|l]=\mathsf{P}\left[\sum_{j=1}^{k}b_1(j)\leq\sum_{j=1}^{l}b_2(j)\text{ and }\sum_{j=1}^{k+1}b_1(j)>\sum_{j=1}^{l}b_2(j)\right]$$

The probability can be rewritten as

$$\mathsf{P}[K=k|l] = \mathsf{P}[0 \leq Z - Y < X]$$

where $X, Y, Z$ are independent random variables. $X = b_1(k+1)$ is exponentially distributed with probability density

$$f_X(x) = \lambda\, e^{-\lambda\, x}.$$

$Y = \sum_{j=1}^{k} b_1(j)$ and $Z = \sum_{j=1}^{l} b_2(j)$ are sums of $k$ respectively $l$ exponentially distributed random variables, that is $Y, Z$ are Gamma (also known as m-Erlang) distributed. The probability density functions of $Y$ and $Z$ are

$$f_Y(y) = \frac{\lambda\, e^{-\lambda\, y}\, (\lambda\, y)^{k-1}}{(k-1)!}$$

$$f_Z(z) = \frac{\lambda\, e^{-\lambda\, z}\, (\lambda\, z)^{l-1}}{(l-1)!}$$

respectively. We calculate the probability $\mathsf{P}[K=k]$ as follows

$$\mathsf{P}[K=k|l] = \iiint_{0 \leq z-y < x} f_Z(z) f_Y(y) f_X(x) dx dy dz$$

$$= \int_{z=0}^{\infty} \int_{y=0}^{z} \int_{x=z-y}^{\infty} f_Z(z) f_Y(y) f_X(x) dx dy dz$$

Solving the first two integrals leads to

$$\mathsf{P}[K=k|l] = \frac{\lambda^{k+l}}{(l-1)!\, k!} \int_{z=0}^{\infty} e^{-2\lambda z}\, z^{k+l-1}\, dz$$

We can solve this equation using integration by parts. After the first step this leads to

$$\mathsf{P}[K=k|l] = \frac{\lambda^{k+l}}{(l-1)!\, k!} \frac{(k+l-1)}{2\lambda} \int_{z=0}^{\infty} e^{-2\lambda z} z^{k+l-2}\, dz$$

We have to repeat the integration by parts another $k+l-2$ times to find

$$\mathsf{P}[K=k|l] = \frac{\lambda^{k+l}}{(l-1)!\, k!} \frac{(k+l-1)!}{(2\lambda)^{k+l-1}} \int_{z=0}^{\infty} e^{-2\lambda z}\, dz.$$

The integral evaluates to $1/2\lambda$ such that

$$\mathsf{P}[K=k|l] = \frac{1}{(l-1)!\, k!} \frac{(k+l-1)!}{2^{k+l}}$$

and after some reordering

$$\mathsf{P}[K=k|l] = 2^{-(k+l)} \binom{k+l-1}{k}$$

becomes negative binomial. ∎

## REFERENCES

[1] V. Bharghavan, A. Demers, S. Shenker, and L. Zhang, "MACAW: a media access protocol for wireless LAN's," in *Proc. of ACM SIGCOMM*, 1994, pp. 212–225.

[2] G. Bianchi, "Performance analysis of the IEEE 802.11 distributed coordination function," *IEEE J. Select. Areas Commun.*, vol. 18, no. 3, pp. 535–547, Mar. 2000.

[3] C. E. Koksal, H. Kassab, and H. Balakrishnan, "An analysis of short-term fairness in wireless media access protocols," in *Proc. of ACM SIGMETRICS*, June 2000, pp. 118–119.

[4] G. Berger-Sabbatel, A. Duda, O. Gaudoin, M. Heusse, and F. Rousseau, "Fairness and its impact on delay in 802.11 networks," in *Proc. of IEEE GLOBECOM*, 2004, pp. 2967–2973.

[5] B.-J. Kwak, N.-O. Song, and L. E. Miller, "Performance analysis of exponential backoff," *IEEE/ACM Trans. Networking*, vol. 13, no. 2, pp. 343–355, 2005.

[6] T. Sakurai and H. L. Vu, "MAC access delay of IEEE 802.11 DCF," *IEEE Trans. Wireless Communications*, vol. 6, no. 5, pp. 1702–1710, May 2007.

[7] A. K. Parekh and R. G. Gallager, "A generalized processor sharing approach to flow control in integrated services networks: The single-node case," *IEEE/ACM Trans. Networking*, vol. 1, no. 3, pp. 344–357, June 1993.

[8] N. H. Vaidya, P. Bahl, and S. Gupta, "Distributed fair scheduling in a wireless lan," in *Proc. of ACM MOBICOM*, Aug. 2000, pp. 167–178.

[9] A. Banchs and X. Perez, "Distributed weighted fair queuing in 802.11 wireless LAN," in *Proc. of IEEE ICC*, May 2002, pp. 3121–3127.

[10] Y. Kwon, Y. Fang, and H. Latchman, "A novel MAC protocol with fast collision resolution for wireless LANs," in *Proc. of IEEE INFOCOM*, Apr. 2003, pp. 853–862.

[11] P. Goyal, S. S. Lam, and H. M. Vin, "Determining end-to-end delay bounds in heterogeneous networks," *Multimedia Systems*, vol. 5, no. 3, pp. 157–163, May 1997.

[12] D. Stiliadis and A. Varma, "Latency-rate servers: A general model for analysis of traffic scheduling algorithms," *IEEE/ACM Trans. Networking*, vol. 6, no. 5, pp. 611–624, Oct. 1998.

[13] Y. Jiang, "Relationship between guaranteed rate server and lateny rate server," *Computer Networks*, vol. 43, no. 3, pp. 307–315, Oct. 2003.

[14] J. C. R. Bennett, K. Benson, W. F. Courtney, and J.-Y. Le Boudec, "Delay jitter bounds and packet scale rate guarantee for expedited forwarding," *IEEE/ACM Trans. Networking*, vol. 10, no. 4, pp. 529–540, Aug. 2002.

[15] R. L. Cruz, "A calculus for network delay, Part I and Part II," *IEEE Trans. Inform. Theory*, vol. 37, no. 1, pp. 114–141, Jan. 1991.

[16] C.-S. Chang, *Performance Guarantees in Communication Networks*. Springer-Verlag, 2000.

[17] J.-Y. Le Boudec and P. Thiran, *Network Calculus A Theory of Deterministic Queuing Systems for the Internet*. Springer-Verlag, 2001.

[18] F. Ciucu, A. Burchard, and J. Liebeherr, "Scaling properties of statistical end-to-end bounds in the network calculus," *IEEE/ACM Trans. Networking*, vol. 14, no. 6, pp. 2300–2312, June 2006.

[19] M. Fidler, "An end-to-end probabilistic network calculus with moment generating functions," in *Proc. of IWQoS*, June 2006, pp. 261–270.

[20] Y. Jiang, "A basic stochastic network calculus," in *Proc. ACM SIGCOMM*, Oct. 2006, pp. 123–134.

[21] D. Wu and R. Negi, "Effective capacity: A wireless link model for support of quality of service," *IEEE Trans. Wireless Communications*, vol. 2, no. 4, pp. 630–643, July 2003.

[22] M. Fidler, "A network calculus approach to probabilistic quality of service analysis of fading channels," in *Proc. of IEEE GLOBECOM*, Nov. 2006.

[23] R. Jain, D.-M. Chiu, and W. R. Hawe, "A quantitative measure of fairness and discrimination for resource allocation in shared computer system," Digital Equipment, Tech. Rep. DEC-TR-301, Sept. 1984.

[24] G. Berger-Sabbatel, A. Duda, M. Heusse, and F. Rousseau, "Short-term fairness of 802.11 networks with several hosts," in *Proc. of IFIP MWCN*, Oct. 2004, pp. 263–274.

[25] L. B. Jiang and S. C. Liew, "Proportional fairness in wireless LANs and ad hoc networks," *IEEE WCNC*, pp. 1551–1556, Mar. 2005.

[26] A. Babu and L. Jacob, "Fairness analysis of IEEE 802.11 multirate wireless LANs," *IEEE Trans. Veh. Technol.*, pp. 3073–3088, Sept. 2007.

[27] T. Joshi, A. Mukherjee, Y. Yoo, and D. P. Agrawal, "Airtime fairness for IEEE 802.11 multirate networks," *IEEE Trans. Mobile Comput.*, pp. 513–527, Apr. 2008.

[28] Y. Jian and S. Chen, "Can CSMA/CA networks be made fair," in *Proc. of ACM MobiCom*, 2008, pp. 137–144.

[29] "IEEE Standard for Wireless LAN Medium Access Control (MAC) and Physical Layer (PHY) Spedifications," June 2007, Std. 802.11-2007.

[30] S. Sharma, "Analysis of 802.11b MAC: A QoS, Fairness, and Performance Perspective, Tech. Rep. arXiv:cs/0411017v1, Nov. 2004.

[31] A. Kumar, D. Manjunath, and J. Kuri, *Communication Networking: An Analytical Approach*. Morgan Kaufmann, 2004.

[32] M. M. Carvalho and J. Garcia-Luna-Aceves, "Delay analysis of IEEE 802.11 single-hop networks," in *Proc. of IEEE ICNP*, Nov. 2003.

[33] C. L. Barrett, M. V. Marathe, D. C. Engelhart, and A. Sivasubramaniam, "Analyzing the short-term fairness of IEEE 802.11 in wireless multi-hop radio networks," in *Proc. of IEEE MASCOTS*, 2002, pp. 137–144.

[34] Z. Li, S. Nandi, and A. K. Gupta, "Modeling the short-term unfairness of IEEE 802.11 in presence of hidden terminals," *Performance Evaluation*, vol. 63, no. 4, pp. 441–462, 2006.

[35] M. Bredel and M. Fidler, "A measurement study of bandwidth estimation in IEEE 802.11g wireless LANs using the DCF," in *Proc. of IFIP Networking*, May 2008, pp. 314–325.

[36] "Omnet++: Network simulator," www.omnetpp.org.

[37] "Rude/Crude: UDP Data Emitter/Collector," rude.sourceforge.net.

[38] "dd-wrt: Linux wireless router," www.dd-wrt.com.

[39] S. Ross, *A First Course in Probability*. Pearson Prentice Hall, 2006.

[40] T. M. Cover and J. A. Thomas, *Elements of Information Theory*. Wiley, 1991.

[41] F. Baccelli, G. Cohen, G. J. Olsder, and J.-P. Quadrat, *Synchronization and Linearity: An Algebra for Discrete Event Systems*. Wiley, 1992.

[42] J.-Y. Le Boudec, "Application of network calculus to guaranteed service networks," *IEEE Trans. Inform. Theory*, vol. 44, no. 3, pp. 1087–1096, May 1998.

[43] B. D. O. Anderson and J. B. Moore, *Optimal Filtering*. Dover, 2005.

[44] "Video Trace Library," http://trace.eas.asu.edu/.

[45] "D-ITG: Internet Traffic Generator," www.grid.unina.it/software/ITG.